\documentclass{article}

\usepackage{booktabs} 
\usepackage{tabularx} 
\usepackage{float}
\usepackage{threeparttable}
\usepackage{amsmath} 
\usepackage{arxiv}
\usepackage{bm}
\usepackage{amsmath,amsfonts,amssymb,bm}
\usepackage{newtxtext}
\usepackage{newtxmath}
\usepackage{xcolor}
\usepackage[utf8]{inputenc} 
\usepackage[T1]{fontenc}    
\usepackage{hyperref}       
\usepackage{url}            
\usepackage{booktabs}       
\usepackage{amsfonts}       
\usepackage{nicefrac}       
\usepackage{microtype}      
\usepackage{cleveref}       
\usepackage{lipsum}         
\usepackage{graphicx}
\usepackage{natbib}
\usepackage{doi}
\usepackage{enumitem}

\title{Large-Ensemble Simulations Reveal Links Between Atmospheric Blocking Frequency and Sea Surface Temperature Variability}

\newif\ifuniqueAffiliation
\usepackage{authblk}

\newbox{\orcid}\sbox{\orcid}{\includegraphics[scale=0.06]{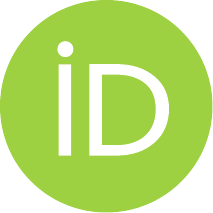}} 
\author[1]{%
	\href{https://orcid.org/0000-0001-5706-592X}{\usebox{\orcid}\hspace{2mm}Zilu Meng\thanks{Correspondence author: Zilu Meng, \texttt{zilumeng@uw.edu}}}%
}
\author[1]{%
	\href{https://orcid.org/0000-0001-8486-9739}{\usebox{\orcid}\hspace{2mm}Gregory J. Hakim}%
}
\author[2]{%
	\href{https://orcid.org/0000-0003-0053-9527}{\usebox{\orcid}\hspace{2mm} Wenchang Yang}%
}
\author[2, 3]{%
	\href{https://orcid.org/0000-0002-5085-224X}{\usebox{\orcid}\hspace{2mm} Gabriel A. Vecchi}%
}
\affil[1]{ 	Department of Atmospheric and Climate Science, University of Washington, Seattle, WA}
\affil[2]{  Department of Geosciences, Princeton University, Princeton, NJ}
\affil[3]{  High Meadows Environmental Institute, Princeton University, Princeton, NJ}

\renewcommand{\headeright}{Meng et al. 2026}
\renewcommand{\undertitle}{}
\renewcommand{\shorttitle}{Large-Ensemble Simulations Reveal Links Between Blocking and SST}

\hypersetup{
colorlinks,
linkcolor={red!50!black},
citecolor={blue!50!black},
urlcolor={blue!80!black}
}

\begin{document}
\maketitle

\begin{abstract}
Atmospheric blocking events drive persistent weather extremes in midlatitudes, but isolating the influence of sea surface temperature (SST) from chaotic internal atmospheric variability on these events remains a challenge. We address this challenge using century-long (1900–2010), large-ensemble simulations with two computationally efficient deep-learning general circulation models. We find these models skillfully reproduce the observed blocking climatology, matching or exceeding the performance of a traditional high-resolution model and representative CMIP6 models. Averaging the large ensembles filters internal atmospheric noise to isolate the SST-forced component of blocking variability, yielding substantially higher correlations with reanalysis than for individual ensemble members. We identify robust teleconnections linking Greenland blocking frequency to North Atlantic SST and El Niño-like patterns. Furthermore, SST-forced trends in blocking frequency show a consistent decline in winter over Greenland, and an increase over Europe. These results demonstrate that SST variability exerts a significant and physically interpretable influence on blocking frequency and establishes large ensembles from deep learning models as a powerful tool for separating forced SST signals from internal noise.
\end{abstract}

\keywords{Atmospheric blocking \and Sea surface temperature \and Deep learning \and Large ensemble \and Climate variability}

\section{Introduction}

Atmospheric blocking events are persistent, quasi-stationary high-pressure systems that interrupt the prevailing midlatitude westerlies, producing extended periods of extreme weather such as cold spells, heatwaves, and droughts \citep[e.g.,][]{hoskins1983shape,holton2013introduction,shuklaSSTAnomaliesBlocking1986,lupoAtmosphericBlockingEvents2021a,sillmann2011extreme}. The dynamics of blocking are inherently complex, involving processes that operate across multiple spatial and temporal scales \citep[e.g.,][]{mullen1987transient,woollings2018blocking,lupoAtmosphericBlockingEvents2021a}. Although numerous theoretical frameworks have been proposed, the mechanisms that control blocking variability and its long-term evolution remain incompletely understood \citep[e.g.,][]{woollings2018blocking,lupoAtmosphericBlockingEvents2021a}.

Blocking events can often be forecast skillfully at the weather timescale, where synoptic-scale dynamics and atmospheric initial conditions provide sufficient predictability \citep{pelly2003well,matsueda2009blocking,nutter1998impact}. However, extending forecasts to the sub-seasonal and seasonal range remains a major challenge \citep{ferranti2018far,matsueda2018estimates}. This difficulty arises because blocking is primarily a high-frequency phenomenon, strongly influenced by transient eddies and internal atmospheric variability that limit the persistence of predictive signals \citep{holton2013introduction,matsueda2009blocking,matsueda2018estimates}. As a result, while numerical weather forecast models can capture the onset and evolution of individual blocking episodes over days to weeks, their skill rapidly diminishes beyond this window. Understanding the sources of predictability for blocking on longer timescales, particularly its potential links to slowly varying boundary conditions such as sea surface temperature (SST) anomalies, remains an open and pressing question \citep[e.g.,][]{lupoAtmosphericBlockingEvents2021a,woollings2018blocking}.

A substantial body of research has examined how the frequency of atmospheric blocking may change in response to global warming. Analyses based on general circulation models (GCMs) generally suggest a decline in blocking occurrence across much of the Northern Hemisphere under future climate scenarios \citep[e.g.,][]{davini2020cmip3,dunn2013northern,woollings2018blocking}. Several studies have linked this reduction to large-scale thermodynamic changes, including the stabilization of the troposphere and modifications of jet stream dynamics \citep{barnes2010influence,matsueda2009future,woollings2018blocking,scaife2010atmospheric}. More recently, attention has turned to the role of SST patterns, with some studies attributing the projected decrease in blocking to the effects of nearly uniform SST warming, which tends to weaken meridional temperature gradients and reduce the persistence of blocking systems \citep{narinesingh2024uniform}. Nonetheless, considerable uncertainty remains, as other modeling studies have highlighted potential regional contrasts—for example, decreases in blocking frequency over the North Pacific but mixed or even opposite trends over the North Atlantic \citep{masato2013winter,dunn2013northern}. This diversity of responses underscores the sensitivity of blocking to both internal variability and the spatial response of warming to external forcing, motivating further investigation into the mechanisms that govern blocking relationships to SST variability. 

Prior research has highlighted the potential influence of large-scale modes of internal climate variability on Northern Hemisphere blocking. For example, the El Niño–Southern Oscillation (ENSO) is known to modulate atmospheric circulation through planetary wave propagation and changes in extratropical storm tracks, thereby affecting the likelihood of blocking episodes in the North Pacific and downstream regions \citep{trenberth1998progress,renwick1996relationships,lupoAtmosphericBlockingEvents2021a}. Similarly, the Pacific Decadal Oscillation (PDO) is associated with basin-wide SST anomalies that alter jet stream behavior and the persistence of midlatitude high-pressure systems \citep{mantua1997pacific,dong2011changes}. Other modes, such as the North Atlantic Oscillation (NAO) and the Arctic Oscillation (AO), are also closely tied to blocking variability, particularly over the Euro-Atlantic sector \citep{woollings2010variability,shabbar2001relationship,lupoAtmosphericBlockingEvents2021a}.

Despite these apparent associations, the robustness of the connection between SST variability and blocking frequency remains limited. Most existing studies rely on observations or reanalysis, and numerical experiments have struggled to establish statistically significant relationships. For instance, \cite{lupoAtmosphericBlockingEvents2021a} note that the relationship between ENSO phase and blocking frequency often fails to exceed the 90\% significance threshold. Similarly, \cite{rohrer2019decadal} use prescribed-SST ensemble experiments from a twentieth-century atmospheric model ensemble (ERA-20CM) \citep{hersbach2015era}, and find no significant differences in blocking frequency between different phases of the AMO or PDO. Earlier studies using reanalysis datasets also find little evidence of systematic blocking changes between El Niño and La Niña winters \citep{wiedenmann2002climatology}. Together, these findings suggest that while SST variability influences large-scale circulation, identifying a direct and robust impact on blocking frequency remains elusive.

Clarifying the causal links between SST anomalies, which provide one of the most important sources of predictability on seasonal to interannual timescales \citep{shuklaSSTAnomaliesBlocking1986, meng2024pacific}, and blocking is challenged by the signal-to-noise problem. A substantial portion of blocking variability arises from sensitivity to atmospheric initial conditions rather than from boundary forcing, which limits the imprint of SST anomalies on blocking frequency \citep{rohrer2019decadal}. In terms of signal-to-noise ratio, this implies that the SST-forced signal is relatively weak and often masked by the large atmospheric internal variability intrinsic to blocking dynamics. 

In this context, isolating the SST-forced signal requires separating it from the influence of initial conditions, which dominate much of the variability in blocking. A standard approach is to conduct large-ensemble simulations in which the same SST boundary conditions are prescribed across many integrations initialized from different atmospheric states \citep{hersbach2015era}. Such an experimental design enables the SST signal to emerge by averaging over internal atmospheric variability. However, recent studies suggest that realistically simulating blocking requires high-resolution general circulation models (GCMs), as coarse-resolution models struggle to capture blocking dynamics with sufficient fidelity \citep{gaoEnhancedSimulationAtmospheric2025a,delucaEnhancedBlockingFrequencies2024}. The high computational cost of such models makes it difficult to perform the large ensemble simulations needed to robustly assess the relationship between SST variability and blocking frequency.

Recent advances in deep learning (DL)–based GCMs demonstrate not only superior skill in short-range weather forecasts compared to traditional models, but also encouraging stability in long-term climate simulations \citep[e.g.,][]{watt-meyerACE2AccuratelyLearning2024, cresswell-clayDeepLearningEarth2024, kochkovNeuralGeneralCirculation2024a,meng2025deep, hua2025extratropical}. These fast emulators provide a new opportunity to perform large ensemble simulations at a fraction of the computational cost of conventional GCMs. We are motivated here by two questions: (i) can deep learning–based GCMs can realistically reproduce blocking frequency?; and (ii) do large ensembles generated with these emulators reveal robust links between SST anomalies and blocking frequency? 

In this study, we employ two deep learning–based GCMs: a fully data-driven model, the Deep Learning Earth System Model (DLESYM) \citep{cresswell-clayDeepLearningEarth2024}, and a hybrid model that couples deep learning components with physical equations, the Neural-GCM (NGCM) \citep{kochkovNeuralGeneralCirculation2024a}. To benchmark their performance, we also include simulations from the traditional physics-based High Resolution Atmospheric Model (HiRAM) \citep{zhaoSimulationsGlobalHurricane2009}. These complementary approaches enable us to address the two central questions outlined above.

\section{Results}

\begin{figure} 
	\centering
	\includegraphics[width=\textwidth]{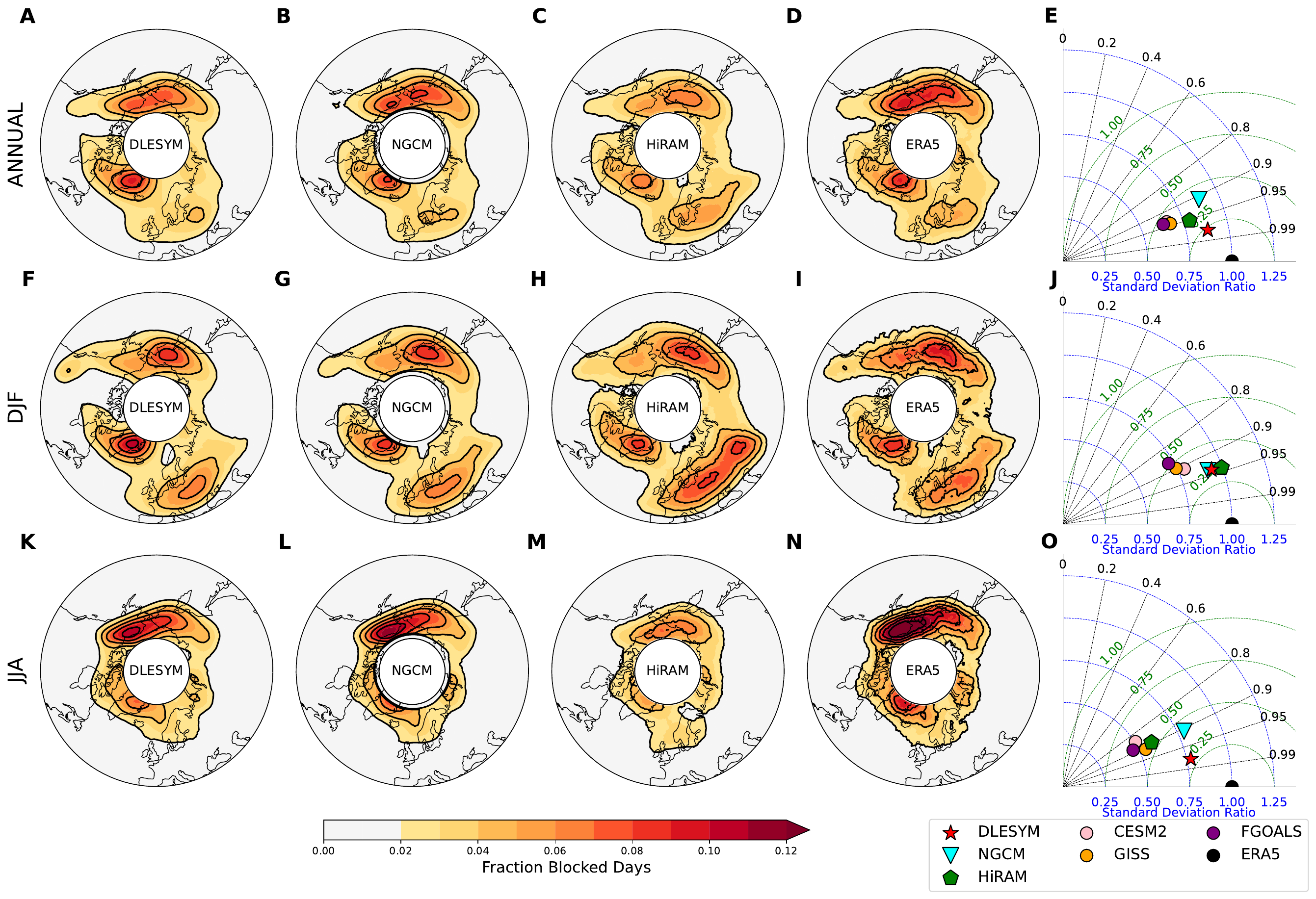}
	\caption{\textbf{Seasonal atmospheric blocking frequency patterns and model evaluation from 1980 to 2010}. Spatial pattern of blocking frequency (fraction of blocked days) for different models (DLESYM, NGCM, HiRAM) and ERA5 reanalysis, shown for (A-D) annual mean, (F-I) December--February (DJF), and (K-N) June--August (JJA). Shading indicates the fraction of blocked days, with contour intervals of 0.2. Panels (E, J, O) show Taylor diagrams comparing model performance against ERA5 in each season, using standard deviation ratios and pattern correlations to quantify spatial agreement, comparing with another 3 CMIP models: CESM2, FGOALS and GISS's AMIP experiments. Models closer to ERA5 indicate better pattern reproduction.}
	\label{fig:pattern}
\end{figure}

We begin by evaluating the ability of the models to reproduce the climatological characteristics of atmospheric blocking given only time-varying surface boundary conditions for SST and sea ice concentration, as in Atmospheric Model Intercomparison Project (AMIP) experiments. This serves as a benchmark for assessing how well the deep learning–based models (DLESYM and NGCM) and the traditional numerical model (HiRAM) compare to observed blocking frequency spatial patterns in the ERA5 reanalysis \citep{hersbachERA5GlobalReanalysis2020}, and with three Coupled Model Intercomparison Project Phase 6 (CMIP6) models (CESM2, FGOALS, and GISS) AMIP experiments\citep{o2016scenario}, from 1980 to 2010. These three models are selected based on their widespread use, data availability, and ability to reasonably simulate large-scale atmospheric circulation \citep{davini2020cmip3,eyring2016overview}, rather than to provide an exhaustive evaluation of all CMIP6 AMIP models. As shown in Figure~\ref{fig:pattern}A–E, for the annual-mean climatology during 1980--2010, DLESYM, NGCM, and HiRAM reproduce the ERA5 pattern more consistently than do the CMIP6 models. HiRAM underestimates the amplitude over North America, while DLESYM provides the closest overall agreement. During DJF (Figure~\ref{fig:pattern}F–J), all three models again capture the ERA5 pattern more realistically than the CMIP6 models, including a more accurate representation of amplitude. In contrast, during JJA (Figure~\ref{fig:pattern}K–O), all models underestimate blocking frequency amplitude, particularly over the North Pacific. Nevertheless, DLESYM and NGCM still outperform both HiRAM and the CMIP6 models in reproducing observed summer blocking patterns. Overall, the two deep learning models exhibit comparable skill, matching or even exceeding the performance of the high-resolution HiRAM while substantially outperforming the lower-resolution CMIP6 models. Previous studies \citep{schiemannNorthernHemisphereBlocking2020a,gaoEnhancedSimulationAtmospheric2025a} have emphasized that higher spatial resolution generally improves the simulation of blocking. Interestingly, that expectation does not hold here: despite operating at coarser resolutions (about 1° for DLESYM and 2.8° for NGCM), both models surpass not only the three CMIP6 models but also the 50-km HiRAM in reproducing blocking patterns. Although the deep learning models are trained on ERA5 data from 1980 to 2020 and thus carry some risk of overfitting, their robust performance represents a significant achievement and highlights the promise of deep learning approaches for advancing the simulation of complex atmospheric variability.

Now we turn to the temporal variability of blocking. A key commonality between AMIP experiments and reanalysis is that both are driven by the same SST and SIC boundary conditions. The distinction, however, is that reanalysis represents a single realization conditioned on observations of the real world, whereas AMIP experiments generate samples of atmospheric states consistent with those boundary conditions. In this context, AMIP large ensembles provide valuable insight on the range of possible atmospheric states constrained by the boundary conditions. In the limit of a large ensemble, the average over internal atmospheric variability independent of the boundary conditions approaches zero, leaving the coupled response to the boundary conditions. 

\begin{figure}
	\centering
	\includegraphics[width=\textwidth]{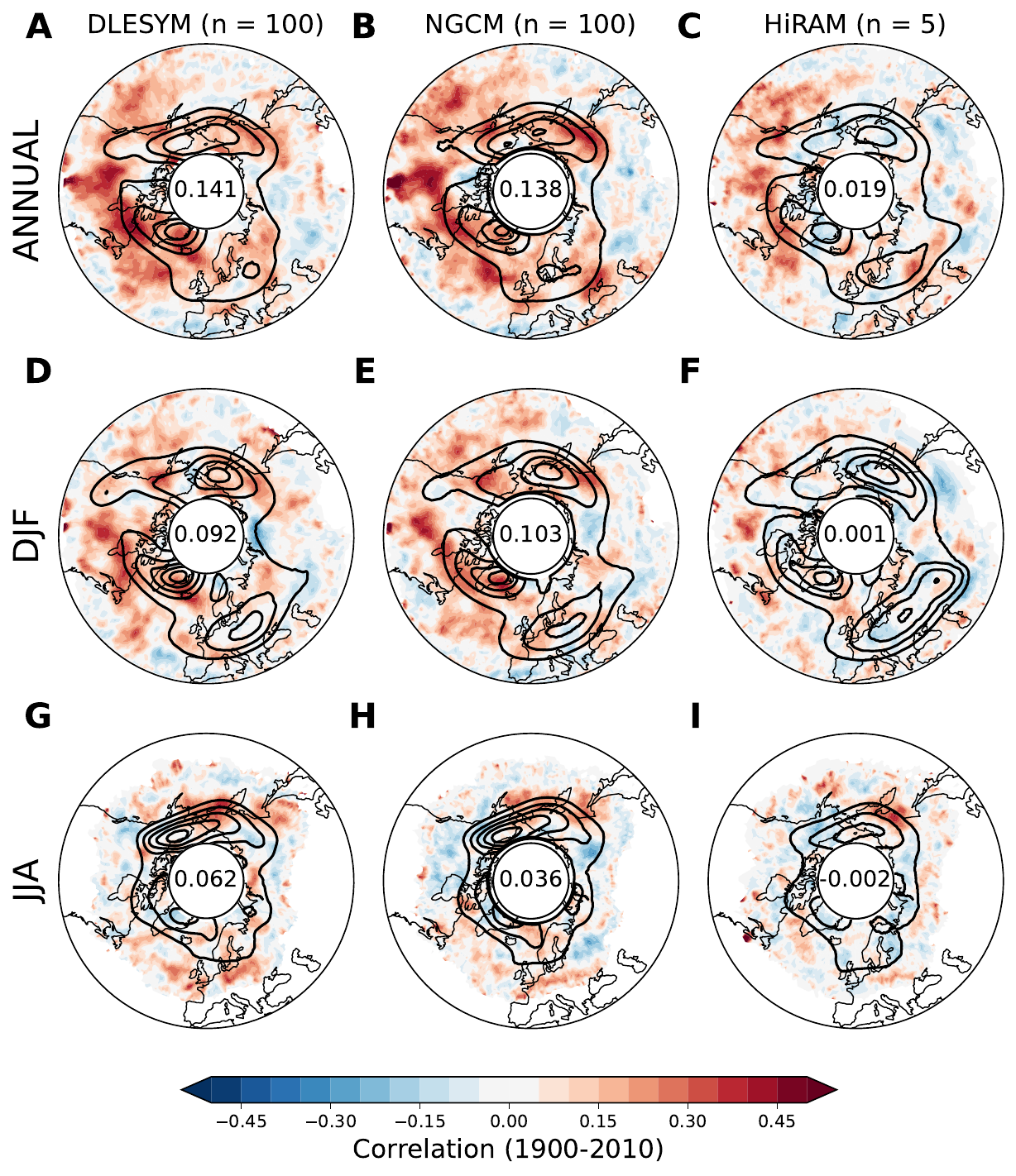}
	\caption{\textbf{Temporal correlation between simulated blocking frequency and reanalysis data}. Spatial distribution of temporal correlations (1900--2010) between simulated blocking frequency ensemble mean and the 20CR reanalysis ensemble mean for (A-C) annual, (D-F) December--February (DJF), and (G-I) June--August (JJA). Results are shown for 100-member DLESYM (A, D, and G), 100-member NGCM (B, E, and H) and 5-member HiRAM (C, F, and I). Shading indicates local correlation coefficients, with warm (cold) colors representing positive (negative) correlations. Black contours represent climatological blocking frequency from each model. Central circle values denote the domain-averaged correlation, weighted by climatological blocking frequency and grid-point area.}
    \label{fig:corr}
\end{figure}

In this study, we use 100 ensemble members from the deep learning models (DLESYM and NGCM) and 5 members from HiRAM to calculate the spatial distribution of temporal correlation between simulated blocking frequencies and those derived from the 20CRv3 reanalysis (Figure~\ref{fig:corr}). Due to small signal-to-noise, the correlations are generally small, but the 100-member DLESYM and NGCM ensembles substantially outperform the 5-member HiRAM ensemble. For the annual mean, DLESYM and NGCM achieve higher correlations than HiRAM, particularly over North America, Greenland, and Europe. DJF results are broadly similar but slightly weaker. In contrast, JJA correlations are much lower than those for the annual mean and DJF, with positive signals largely confined to the Bering Sea and Alaska. The reduced JJA skill may reflect two factors. First, blocking events are generally weaker in summer than in winter, reducing the signal-to-noise ratio and hence the correlation with SST forcing. Second, during DJF, the SST anomalies more effectively trigger deep convection in the tropics (like ENSO and Atlantic Niño \citep{nnamchi2015thermodynamic}), which in turn excites Rossby wave trains \citep{wallace1981teleconnections} that propagate into the midlatitudes and influence blocking activity. This dynamical linkage is weaker in JJA, especially for the extratropical Rossby-wave guide, leading to reduced correlations.

Given the overlap with the 1980-2020 training period for NGCM and DLESYM, we assess the potential for overfitting in the blocking statistics. As shown in Figures~\ref{sfig:corr1} and \ref{sfig:corr2}, the domain-weighted annual mean correlations are comparable between the out-of-sample period (1900–1960) and the in-sample period (1960–2010). Specifically, for DLESYM the correlations are 0.152 (1900--1960) and 0.115 (1960--2010), while for NGCM they are 0.115 (1900--1960) and 0.170 (1960--2010). These results suggest that overfitting is not primarily responsible for the blocking frequency correlation results, particularly for DLESYM, which shows similar or even higher skill in the out-of-sample period. The  increase in correlation for DLESYM and NGCM derives from their computational economy, which facilitates the large ensembles needed to average over internal atmospheric variability, thereby isolating the SST-forced signal, which is obscured in the 5-member HiRAM ensemble. We diagnose this further by bootstrap sampling the DLESYM and NGCM results for 5-member ensembles to illustrate the increase in error for smaller ensembles (Supplementary Figure \ref{sfig:bootstrap}).

\begin{figure}
	\centering
	\includegraphics[width=\textwidth]{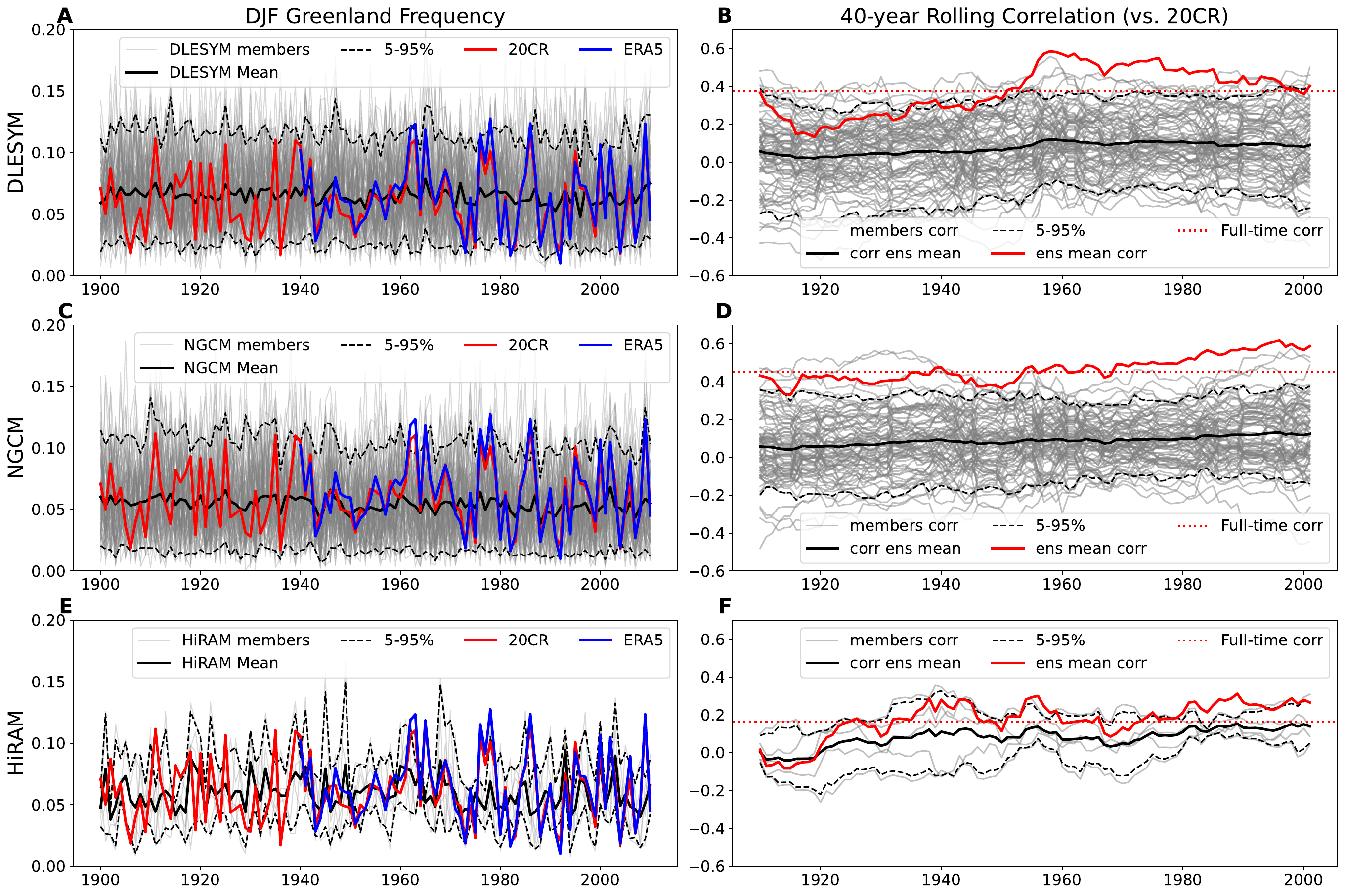}
	\caption{\textbf{Greenland blocking frequency and temporal correlation with reanalysis}. (A, C, E) Time series of DJF Greenland blocking frequency from DLESYM (A), NGCM (C), and HiRAM (E) models, compared with 20CR (red) and ERA5 (blue). Thin gray lines show individual ensemble members; black lines indicate ensemble means. Dashed black lines represent the 5--95\% ensemble spread. (B, D, F) 40-year rolling correlation between model blocking frequency and 20CR. Thin gray lines show correlations from individual ensemble members; the black line shows the ensemble-mean correlation. Dashed lines mark the 5--95\% spread across members. The red line shows the rolling correlation of the ensemble mean against 20CR. Horizontal dotted red lines indicate full-period correlations of ensemble means.}
    \label{fig:greenlandts}
\end{figure}

To provide a more detailed perspective on how simulated blocking compares with reanalysis, we focus here on Greenland DJF blocking. We choose Greenland because blocking in this region has long been a subject of research \citep{rohrer2019decadal,wachowicz2021historical} and is known to exert a significant influence on the Greenland ice sheet \citep{mcleod2016linking}. A prior study \citep{rohrer2019decadal} using 10-member ERA-20CM AMIP experiments finds no correlation between the ensemble mean and reanalysis data, suggesting little connection between observed SST atmospheric variability. Here we revisit this question using 100 member ensembles for DLESYM and NGCM. As shown in Figure~\ref{fig:greenlandts}A–D, when considering the individual ensemble members, it is difficult to  identify clear correlations with reanalysis (20CRv3 and ERA5). However, nearly all of the 20CRv3 values lie within the 5–95\% ensemble spread, indicating that the large ensembles capture the range of observed variability. By contrast, the 5-member HiRAM ensemble fails to encompass the observed variability within its spread. Averaging over the ensemble members filters out much of the internal atmospheric variability that is independent of the surface boundary conditions, isolating the component  linked to SST forcing. Although the amplitude of variability in the ensemble mean is much smaller than that of 20CRv3, the correlation of about 0.4 over the full period suggests that the large-ensemble mean captures part of the timing of observed variability if not the amplitude. In contrast, the HiRAM ensemble mean shows weaker skill, with correlations below 0.2. There is also notable multidecadal variability in the correlations from DLESYM and NGCM (Figure \ref{fig:corr}B and D), but given overlap with the model training period, it is difficult to determine whether this arises from model overfitting or from genuine decadal-scale variability in the climate system. 

To further investigate the pattern of SST associated with Greenland blocking variability, we calculate the temporal correlation with ensemble-mean Greenland blocking frequencies. The results from DLESYM, NGCM, and 20CR exhibit a consistent and physically interpretable pattern (Figure \ref{fig:sstcorr}, with  a North Atlantic tripole SST structure associated with the negative phase of the North Atlantic Oscillation (NAO) \citep{czaja2002observed}. This pattern is characterized by positive SST anomalies in the subpolar North Atlantic near Greenland, negative anomalies in the midlatitudes, and positive anomalies in the tropics. Such a tripole structure is commonly linked to weakened westerlies and enhanced ridging over Greenland, which favor more frequent blocking episodes through the development of anomalously high geopotential heights at 500 hPa \citep{woollings2010variability}. The results also show  an El Niño–like SST anomaly pattern in the tropical Pacific. During El Niño events, enhanced deep convection over the central and eastern Pacific excites large-scale Rossby wave trains that propagate into the extratropics \citep{renwick1996relationships,wallace1981teleconnections}, modifying the North Atlantic jet stream, favoring a negative NAO phase, and thereby increasing the likelihood of Greenland blocking \citep{moron2003seasonal,woollings2010variability}. By contrast, HiRAM does not exhibit such significant SST patterns, likely because the strong atmospheric internal variability overwhelms the boundary-forced signal for the small ensemble (only five members). The influence of El Niño on blocking frequency is further assessed using linear regression onto the Niño-3.4 index, which reveals a consistent ENSO-related response across all models and reanalyses in both winter and the annual mean (Supplementary Fig. \ref{sfig:ninoreg}).

Taken together, these results demonstrate that large ensemble simulations from the deep learning models (DLESYM and NGCM) not only reproduce the observed climatological spatial patterns of blocking but also capture meaningful aspects of its temporal variability. Compared with the smaller ensemble member size for HiRAM, the DL-based ensembles achieve substantially higher correlations with reanalysis data, particularly in winter and over key regions such as Greenland, where the ensemble mean isolates SST-forced signals from internal atmospheric noise. In other words, the larger ensemble enabled by the DL models reveals the component of blocking frequency variability that is driven by SST forcing.

We next use the large ensembles to examine trends in blocking frequency related to SST variability over over the past century (Figure \ref{fig:trend}). While previous studies suggest a decline in blocking under global warming \citep{narinesingh2024uniform, davini2020cmip3, woollings2018blocking}, our analysis extends this perspective by assessing the range in modeled trends in blocking in response to past SST variability. For DJF (Figure \ref{fig:trend}E–H), all three GCMs (DLESYM, NGCM, and HiRAM) show a broadly similar pattern: a decrease in blocking frequency around Greenland and an increase over Europe and central Russia, although the signal is less significant in HiRAM. By contrast, the 20CRv3 reanalysis exhibits weaker and largely insignificant trends across most regions. During JJA (Figure \ref{fig:trend}I–L), the three GCMs agree on a general increase over Greenland and northern Europe, whereas 20CRv3 indicates a more spatially extensive and significant increase. This discrepancy may reflect the fact that summer blocking is more strongly controlled by internal atmospheric variability rather than SST forcing, leading to weaker consistency between models and observations. For the annual mean (Figure \ref{fig:trend}A–D), the patterns are dominated by the DJF signal, with model results again showing reduced Greenland blocking and enhanced European blocking. However, substantial discrepancies remain between the models and 20CRv3, except over Europe where the agreement is stronger. Importantly, despite their very different architectures, DLESYM, NGCM, and HiRAM all produce a consistent annual mean blocking trend over the last century, underscoring the robustness and significance of this simulated pattern. The divergence between the consistent SST-forced trends in our models and the weaker, noisier trend in reanalysis could be due to random internal variability; that is, the observed trend reflects this variability rather than a forced response from changing SSTs.

\section{Conclusion and Discussion}

\begin{figure}
	\centering
	\includegraphics[width=\textwidth]{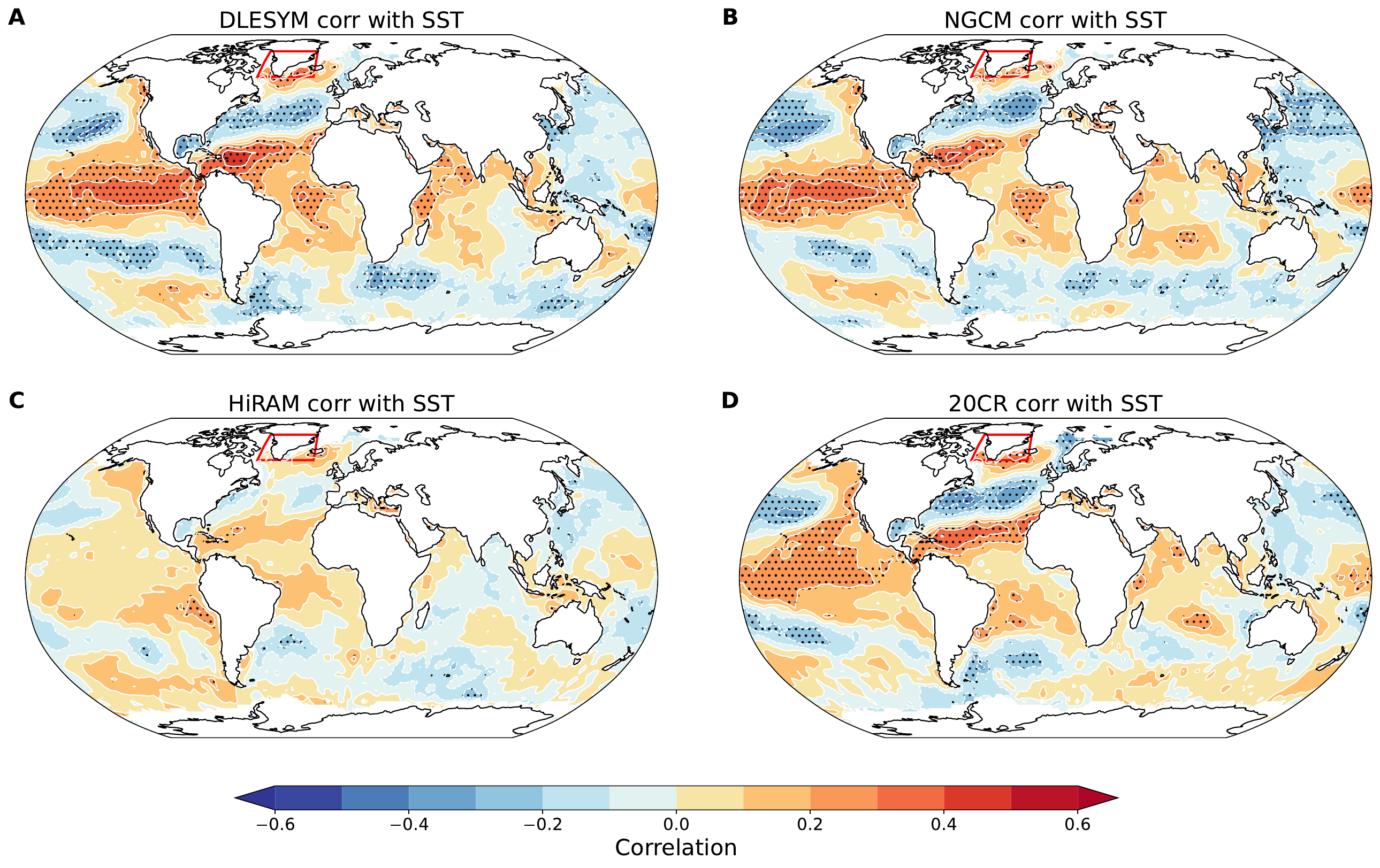}
	\caption{\textbf{Correlation between Greenland blocking frequency and global sea surface temperatures (SSTs)}. Spatial correlation patterns between DJF Greenland blocking frequency ensemble mean and detrended SST anomalies during 1900--2010 for (A) DLESYM, (B) NGCM, (C) HiRAM, and (D) 20CR reanalysis. Colors indicate correlation coefficients, with warm (cool) shading showing regions where blocking frequency increases with warmer (cooler) SSTs. Stippling marks regions where correlations are statistically significant ($p < 0.01$).}
    \label{fig:sstcorr}
\end{figure}

This study leveraged large-ensemble AMIP-style simulations from two deep-learning–based general circulation models (DLESYM and NGCM) and a traditional high-resolution model (HiRAM), together with two reanalyses (ERA5 and 20CRv3), to interrogate the links between SST variability and Northern Hemisphere atmospheric blocking. Three main conclusions emerge. First, the DL-based ensembles reproduce the observed climatology of blocking more faithfully than three representative CMIP6 AMIP model simulations and match or exceed the performance of the high-resolution HiRAM, despite operating at coarser grid resolution. Second, large ensembles are essential for extracting the boundary-forced (SST, SIC) signal: averaging across 100 members in DLESYM and NGCM isolates SST-related variability and yields substantially higher correlations with reanalysis than the 5-member HiRAM ensemble, particularly in DJF blocking frequency over the North Atlantic. Third, we recover consistent, physically interpretable teleconnections for Greenland blocking: an Atlantic tripole SST pattern linked to negative NAO and an El Niño–like tropical Pacific pattern. The consistency between these results and reanalysis teleconnections increases confidence that these DL models capture physically consistent SST teleconnection patterns as drivers of Greenland blocking (Figure~\ref{fig:sstcorr}). This suggests that the models have learned dynamically meaningful relationships from the training data.

A further concern is the ensemble size necessary to reliably isolate the boundary-forced signal arising from SST and SIC variability. Using a bootstrap method, we find that the area-climatology-weighted mean correlation skill over the Northern Hemisphere increases monotonically with ensemble size and converges to a quasi-stable level near 0.13 (Fig. \ref{fig:corrwithsize}), suggesting that this ensemble size is sufficient to capture the dominant boundary-forced signal.

\begin{figure}
    \centering
    \includegraphics[width=\textwidth]{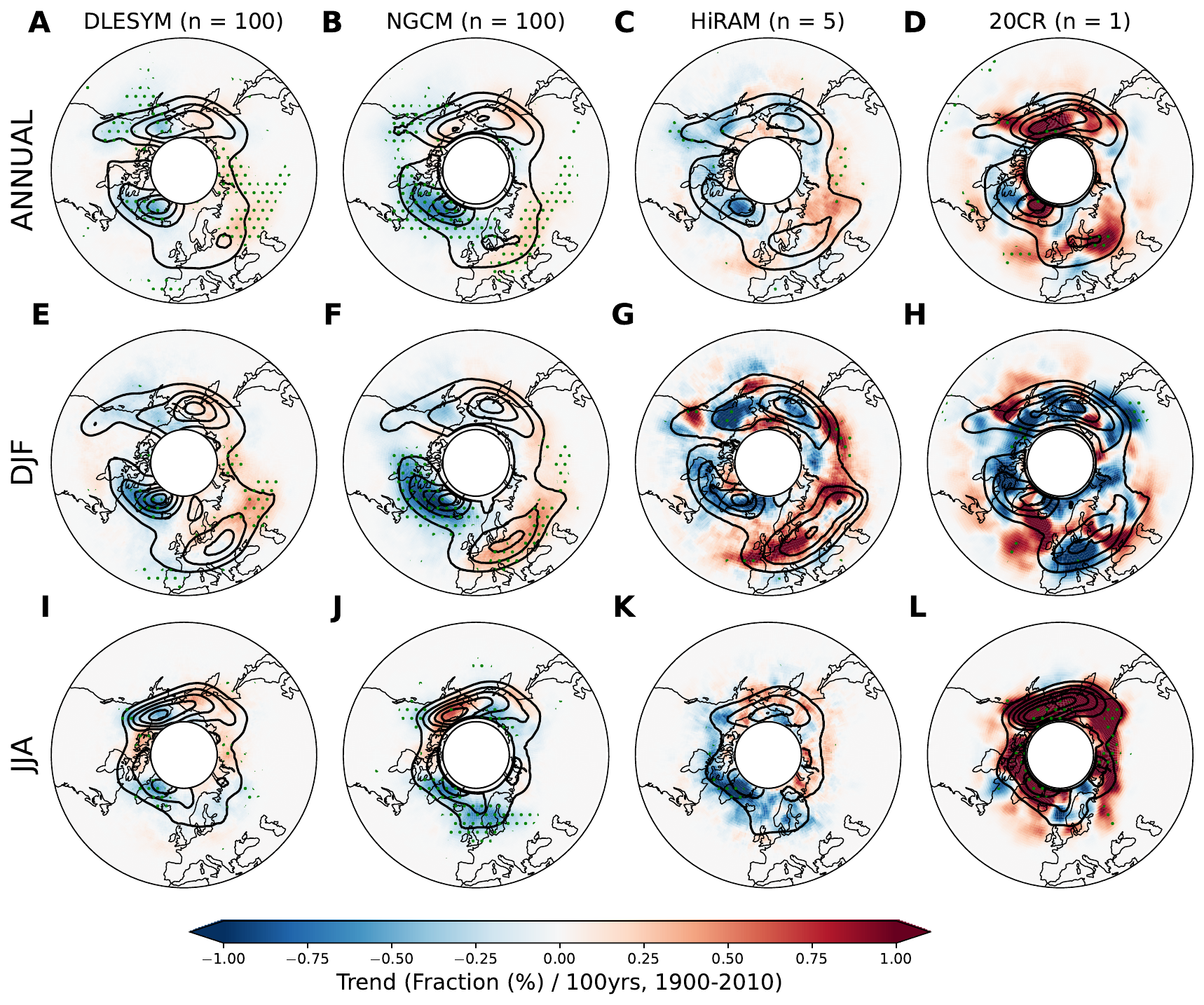}
    \caption{\textbf{Trends in blocking frequency from 1900 to 2010}. Spatial distribution of linear trends in ensemble-mean blocking frequency (\% per 100 years) for (A-D) annual, (E-H) DJF, and (I-L) JJA seasons. Results are shown for DLESYM (n = 100), NGCM (n = 100), HiRAM (n = 5), and 20CR (n = 1). Shading indicates the magnitude and sign of the trends, with red (blue) areas representing increasing (decreasing) blocking frequencies. Black contours show climatological blocking frequency patterns. Green stippling marks areas where trends are statistically significant ($p < 0.01$).}
    \label{fig:trend}
\end{figure}

Long-term trends derived from the ensembles (Figure~\ref{fig:trend}) provide an SST-forced perspective on centennial-scale changes in blocking. In DJF, the three models consistently indicate reduced Greenland blocking and increased European blocking, whereas 20CRv3 shows weaker or insignificant trends at many locations. This model--reanalysis discrepancy likely reflects differences in what is being diagnosed: our ensembles isolate the response to observed SSTs, while reanalysis integrates all forcings (including non-SST radiative forcings) and internal variability into the single observed realization. Notably, the agreement among DLESYM, NGCM, and HiRAM—despite their very different architectures—suggests that these trends represent a robust SST-forced signal.

In summary, our results show that (i) large ensembles are very helpful for extracting the SST-forced component of blocking variability; (ii) DL-based GCMs can reproduce observed blocking climatology and blocking teleconnection patterns with skill comparable to or exceeding that of a high-resolution traditional model; and (iii) Some SST patterns provide consistent, mechanistically plausible links to blocking on seasonal to multidecadal timescales. Together, these findings advance the case that SST variability exerts a detectable and interpretable influence on NH blocking and demonstrate how deep learning–enabled ensembles can sharpen the separation between boundary-forced signals and internal atmospheric noise. Future work may address some of the limitations of this study, including larger ensembles from a traditional model, different forcings (e.g. aerosols and Greenhouse gases), and diagnostics aimed at isolating mechanisms for SST--blocking connections. These results open a pathway toward a new generation of climate experiments in which the predictability, forcing sensitivity, and mechanistic underpinnings of atmospheric extremes can be systematically explored using deep learning–enabled large ensembles.

\section{Models, Data and Methods}

\subsection{Models}

We utilize three types of general circulation models: a purely data-driven DL-based GCM, DLESYM \citep{cresswell-clayDeepLearningEarth2024}; a physics–DL hybrid model, NGCM \citep{kochkovNeuralGeneralCirculation2024a}; and a traditional physics-based atmospheric GCM, HiRAM \citep{zhaoSimulationsGlobalHurricane2009}. 

\subsubsection{Deep Learning Earth SYstem Model (DLESYM)}

The DLESYM is a purely data-driven DL-based GCM \citep{cresswell-clayDeepLearningEarth2024}. The model uses a U-net architecture \citep{ronneberger2015u} with HEALPix mesh \citep{gorski2005healpix} that takes as input the current state $\mathbf{x}_t$ and 6-hour-earlier state $\mathbf{x}_{t-6hr}$ to predict the next state 6-hours later $\mathbf{x}_{t+6hr}$. The DLESYM is a ocean--atmosphere coupled model, but only the atmosphere component is predicted for this study as discussed below. The spatial resolution is approximately 1° × 1°, achieved by dividing the 12 primary HEALPix faces into $64 \times 64$ cells, resulting in a total of 49,152 grid cells globally \citep{gorski2005healpix}. The atmospheric model is trained using ERA5 reanalysis data \citep{hersbachERA5GlobalReanalysis2020}, and outgoing longwave radiation (OLR) observational data from the International Satellite Cloud Climatology Project (ISCCP) \citep{rossow2004international} for the period 1983––2016. 

\subsubsection{Neural General Circulation Model (NGCM)}

The Neural General Circulation Model (NGCM) is a physics--DL hybrid model, which makes forecasts by combining physical equations with a deep learning parameterization, as represented by \cite{kochkovNeuralGeneralCirculation2024a}:
\begin{equation}
    \frac{\partial \mathbf{x}}{\partial t} = \Phi(\mathbf{x}) + \Psi(\mathbf{x}).
\end{equation}
\noindent Here $\mathbf{x}$ denotes the state vector (vorticity, divergence, temperature anomaly, and log of surface pressure), $\Phi$ represents the tendency from physical conservation laws (e.g., vorticity and divergence tendency equations), and $\Psi$ is a deep learning model that has been trained on the residual between solutions of the physical equations and the ERA5 reanalysis dataset \citep{hersbachERA5GlobalReanalysis2020} during 1979--2017. In this study, we use the NGCM-2.8° deterministic model, which has a spatial resolution of approximately 2.8° and shows stable climate simulations as discussed in \cite{kochkovNeuralGeneralCirculation2024a}. To ensure numerical stability in multi-century integrations, the global-mean surface pressure is fixed to a constant value, following the methodology described in \cite{neuralgcm2024}.

\subsubsection{High-Resolution Atmospheric Model (HiRAM)}

The Geophysical Fluid Dynamics Laboratory (GFDL) High-Resolution Atmospheric Model (HiRAM) \citep{zhaoSimulationsGlobalHurricane2009} is a traditional  physical atmospheric GCM developed at GFDL. HiRAM numerically solves the governing equations of atmospheric dynamics and thermodynamics, supplemented by parameterization schemes to represent unresolved processes such as convection, radiation, and cloud microphysics. At the horizontal grid spacing considered here (around 50km), which is finer than that of typical climate models, HiRAM is capable of simulating the statistics of small-scale features such as tropical cyclones, as demonstrated by \cite{zhaoSimulationsGlobalHurricane2009}, \cite{harris2016high}, \cite{Chan2021}, and \cite{Yang2021}.

\subsubsection{Computational expense}

As discussed in the introduction, a key consideration for conducting large-ensemble simulations is the computational cost of running the models. Table \ref{tab:runtime} summarizes the required hardware, runtime, and spatial resolution for each model. The two deep learning–based models complete 100-year simulations in a fraction of the time and with far fewer CPUs compared to the traditional numerical model HiRAM. This efficiency makes it feasible to generate the large ensembles needed to robustly investigate the connection between blocking frequencies and SST variability. DLESYM is faster than NGCM because it advances the model state using a U-Net–based inference step to predict the next time step, whereas NGCM requires computationally expensive numerical integration of the governing equations in addition to its neural parameterization.

\begin{table}[ht]
\centering
\caption{Computational resources and runtime for each model per 100 years.}
\begin{tabular}{lccc}
\hline
\textbf{Model} & \textbf{DLESYM} & \textbf{NGCM} & \textbf{HiRAM} \\
\hline
Device & 1 A100 GPU, 1 CPU & 1 A100 GPU, 1 CPU & 540 CPUs \\
Runtime   & 1.2 hours & 6 hours & 200 hours \\
Spatial Resolution & around 1°& around 2.8° & around 50km \\
\hline
\end{tabular}
\label{tab:runtime}
\end{table}

\subsection{Experimental Design}

We conduct experiments following the Atmospheric Model Intercomparison Project (AMIP) protocol \citep{gates1999overview,eyring2016overview,meng2024pacific} using NGCM, DLESYM, and HiRAM. All simulations are forced with observed sea surface temperature (SST) and sea-ice-concentration (SIC) boundary conditions from 1900 to 2020, based on the HadISST dataset \citep{rayner2003global}. This period allows us to assess model generalization: since NGCM and DLESYM are trained on ERA5 reanalysis for 1980–-2020, the years 1900--1960 serve as an out-of-sample test period, while 1961--2010 represent the in-sample period. DLESYM is driven exclusively by SST boundary conditions, whereas NGCM and HiRAM are forced with both SST and SIC. For radiative forcings other than volcanic aerosols, HiRAM follows the CMIP5 historical scenario through 2005 and RCP4.5 thereafter \citep{taylor2012overview}. Volcanic forcings are prescribed from the CMIP6 historical scenario through 2014 \citep{o2016scenario} and SSP2-4.5 thereafter, reflecting substantial differences between CMIP5 and CMIP6 reconstructions \citep{Yang2019,Jacobson2020}. Initial conditions for NGCM and DLESYM are randomly sampled from ERA5 states between 1980 and 2020. We generate 100 ensemble members for both NGCM and DLESYM, and 5 ensemble members for HiRAM.

\subsection{Data}

The 20th Century Reanalysis Version 3 (20CRv3) \citep{slivinski2021evaluation} provides global atmospheric reanalysis data spanning 1836--2015. It is produced by assimilating surface pressure observations over land and ocean using an ensemble Kalman filter, with prescribed SST and sea ice concentration (SIC) as boundary conditions. In this study, we use all 80 ensemble members of 20CRv3 to calculate blocking frequency and take the ensemble mean as the reference. For additional verification, we also use ERA5 reanalysis \citep{hersbachERA5GlobalReanalysis2020}, which extends from 1940 to the present and assimilates a broader range of satellite and in situ observations. The blocking frequencies derived from 20CRv3 and ERA5 are highly correlated over their overlapping period (1940---2010, figure not shown). Consequently, 20CRv3 serves as our primary reference for evaluating model simulations, while ERA5 is used to assess blocking patterns during the satellite era (1980--2010).

To further assess the simulation skill of the two deep learning models (NGCM and DLESYM), we compare their blocking frequency patterns against those from the AMIP experiments \citep{o2016scenario} of three CMIP6 models--CESM2, GISS, and FGOALS--over the period 1980--2010.

\subsection{Blocking Definition}

The choice of a blocking index is subjective and therefore represents an unavoidable source of uncertainty \citep{wachowicz2021historical,davini2020cmip3,lupoAtmosphericBlockingEvents2021a}. Blocking indices are generally grouped into three main categories: (i) flow-topology indices, which identify reversals in geopotential height or potential vorticity (PV) gradients \citep{tibaldi1990operational, davini2020cmip3,pelly2003new}; (ii) anomaly-based indices, which detect departures relative to a climatological mean or threshold \citep{dole1983persistent,schwierz2004perspicacious}; and (iii) hybrid indices, which combine aspects of both approaches \citep{dunn2013northern,woollings2018blocking}. While all of these approaches capture the primary features of blocking climatology, they may yield differences in variability and trends depending on the region, season and time periods considered.

In this work, we adopt a two-dimensional (2D) blocking index based on meridional gradients of the 500hPa geopotential height field  \citep{davini2020cmip3}, which belongs to the first category. A gradient-based index is preferred over anomaly-based methods because the latter are sensitive to the choice of baseline climatology. This issue becomes particularly problematic when assessing blocking under climate change, where shifts in the mean state can bias anomaly-based detection. For example, under global warming the 500hPa geopotential height exhibits an overall increase \citep{santerContributionsAnthropogenicNatural2003}; consequently, anomaly-based methods would artificially indicate a rising trend in blocking frequency simply due to the background warming \citep{wachowicz2021historical}. We further opt for the 500hPa geopotential height field, rather than potential vorticity, because the DLESYM model output does not include the variables required for potential vorticity calculations \citep{holton2013introduction}. The index is defined by three meridional gradients:  
\begin{equation}
\text{GHGS}(\lambda_{0}, \phi_{0}) =
\frac{Z500(\lambda_{0}, \phi_{0}) - Z500(\lambda_{0}, \phi_{S})}{\phi_{0} - \phi_{S}},
\end{equation}
\begin{equation}
\text{GHGN}(\lambda_{0}, \phi_{0}) =
\frac{Z500(\lambda_{0}, \phi_{N}) - Z500(\lambda_{0}, \phi_{0})}{\phi_{N} - \phi_{0}},
\end{equation}
\begin{equation}
\text{GHGS}_{2}(\lambda_{0}, \phi_{0}) =
\frac{Z500(\lambda_{0}, \phi_{S}) - Z500(\lambda_{0}, \phi_{S_{2}})}{\phi_{S} - \phi_{S_{2}}},
\end{equation}
\noindent where $\phi_{0}$ denotes latitude (30°N–75°N) and $\lambda_{0}$ denotes longitude (0°–360°). The reference latitudes are defined as $\phi_{S} = \phi_{0} - 15^{\circ}$, $\phi_{N} = \phi_{0} + 15^{\circ}$, and $\phi_{S_{2}} = \phi_{0} - 30^{\circ}$.  

A grid point is classified as blocked if the following conditions are satisfied simultaneously:  
\begin{equation}
\text{GHGS}(\lambda_{0}, \phi_{0}) > 0, \quad
\text{GHGN}(\lambda_{0}, \phi_{0}) < -\frac{10 \, \text{m}}{^\circ \text{lat}}, \quad
\text{GHGS}_{2}(\lambda_{0}, \phi_{0}) < -\frac{5 \, \text{m}}{^\circ \text{lat}}.
\end{equation}

These conditions correspond to an anomalous easterly flow in the midlatitudes, flanked by westerly flow to both the north and the south of the blocked grid point. Blocking frequency is then quantified as the percentage of days classified as blocked within a given season (DJF, JJA, and annual in this study).

It is important to note that we do not impose additional temporal or spatial filtering, such as requiring a minimum duration of five consecutive days for blocking events \citep{rohrer2019decadal}. The decision is motivated by three considerations: (i) Previous studies \citep{davini2020cmip3,davini2012bidimensional} have shown that results with and without persistence filtering are highly correlated; (ii) A substantial body of prior work has adopted this simplified definition, including studies such as  \citep{tibaldi1990operational,davini2020cmip3,davini2012bidimensional,prodhomme2016benefits}; (iii) The approach is straightforward to apply and computationally efficient, which is an important consideration for  large ensembles.


\clearpage 

%

%
%
%
%
%
%


\section*{Acknowledgments}
We acknowledge the conversations with Nathaniel Cresswell-Clay (University of Washington), Dale Durran (University of Washington), Oliver Watt-Meyer (AI2 company), Dmitrii Kochkov (Google Research), Vince Cooper (University of Washington), and Dominik Stiller (University of Washington) on the model configuration and AMIP experiments running. Z.M. also acknowledges the conversations with Zhanxiang Hua (University of Washington) and Zhaoyu Liu (Purdue University) about blocking index calculations. 


\paragraph*{Funding:}
This research was supported by NSF awards 2501400, 2530556, 2402475, 2202526 and 2105805, and Heising-Simons Foundation award 2023-4715.

\paragraph*{Competing interests:}
There are no competing interests to declare.

\paragraph*{Data and materials availability:}
The plotting package SACPY \citep{meng2021research,meng2023sacpy,meng2025coupled} is located at \url{https://github.com/ZiluM/sacpy}. The NGCM is located at \url{https://neuralgcm.readthedocs.io/en/latest/} and the DL\textit{ESy}M is located at \url{https://github.com/AtmosSci-DLESM/DLESyM}. The 20CRv3 dataset is located at \url{https://www.esrl.noaa.gov/psd/data/20thC_Rean/}. The HadISST dataset is located at \url{https://www.metoffice.gov.uk/hadobs/hadisst/}. The ERA5 is located at \url{https://doi.org/10.24381/cds.adbb2d47}. 

\clearpage

\section*{Supplementary Information}

\setcounter{figure}{0}
\renewcommand{\thefigure}{S\arabic{figure}}

\begin{figure}
    \centering
    \includegraphics[width=\linewidth]{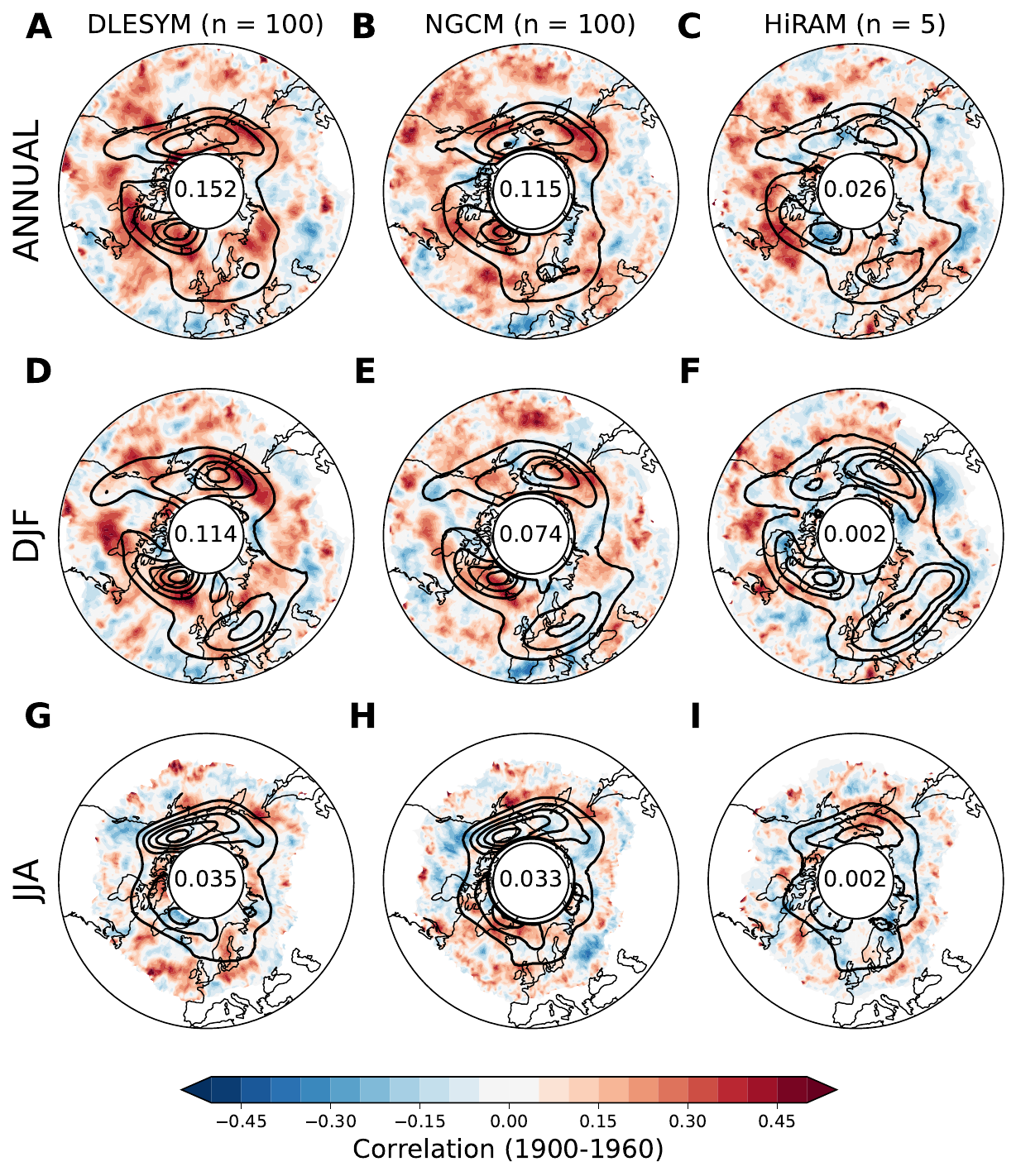}
    \caption{Same as Figure \ref{fig:corr}, but for the period from 1900 to 1960.}
    \label{sfig:corr1}
\end{figure}

\begin{figure}
    \centering
    \includegraphics[width=\linewidth]{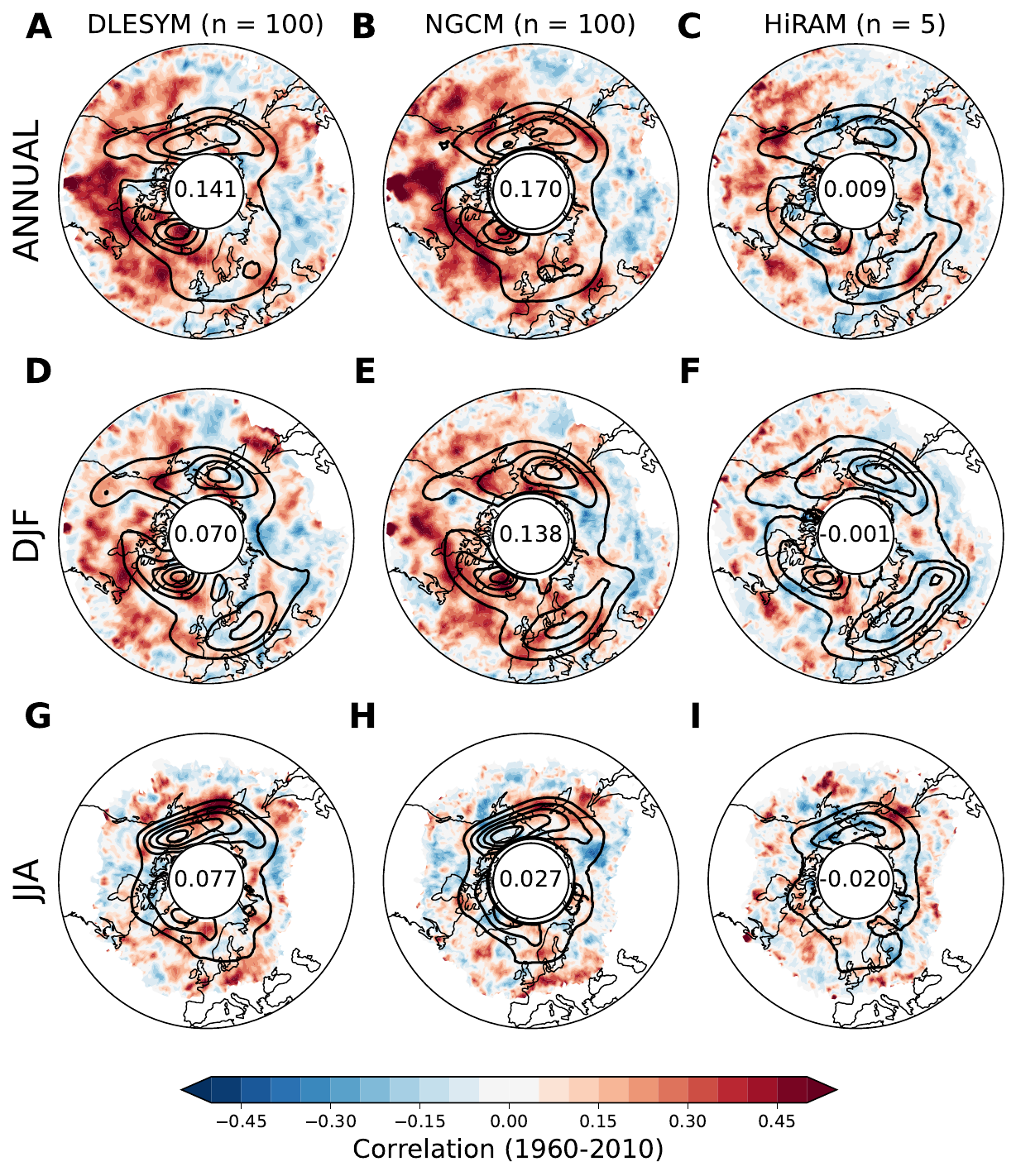}
    \caption{Same as Figure \ref{fig:corr}, but for the period from 1960 to 2010.}
    \label{sfig:corr2}
\end{figure}

\begin{figure}
    \centering
    \includegraphics[width=1\linewidth]{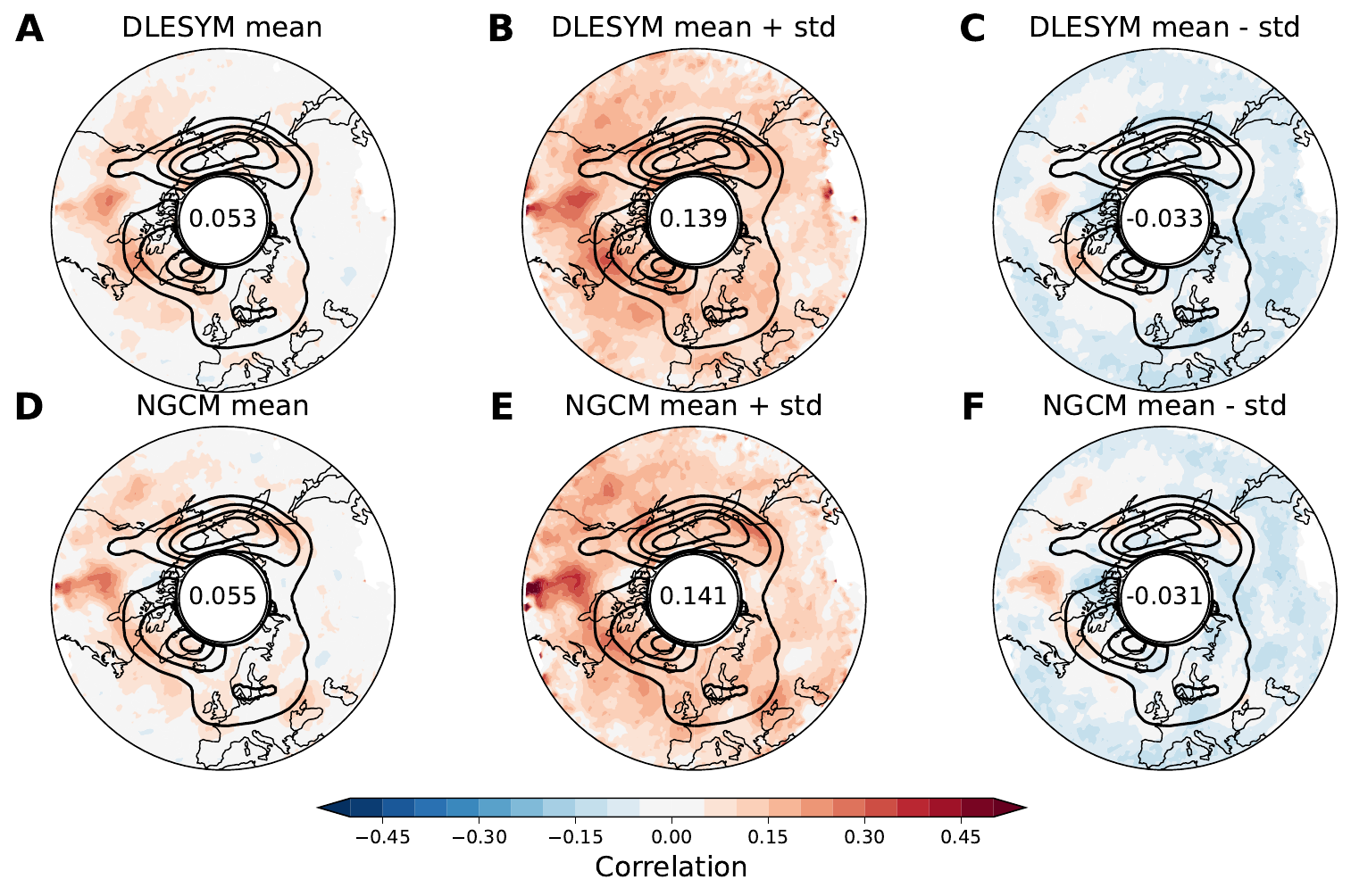}
    \caption{Spatial patterns of temporal correlation (1900--2010) between simulated and observed blocking frequency based on a five-member bootstrap strategy. Panels (A–C) show results for DLESyM and panels (D–F) for NGCM. The left column (A, D) presents the mean correlation maps obtained by averaging over all 1000 bootstrap realizations, where each realization is constructed by randomly selecting five ensemble members to mimic the ensemble size of HiRAM. The middle (B, E) and right (C, F) columns show the mean plus and minus one standard deviation of the bootstrap correlation distributions, respectively, illustrating the sensitivity of the correlations to ensemble sampling. Black contours indicate the climatological mean blocking frequency from 20CR. The numbers in the center of each panel denote the area-climatology-weighted mean correlation over the plotted domain.}
    \label{sfig:bootstrap}
\end{figure}

\begin{figure}
    \centering
    \includegraphics[width=\linewidth]{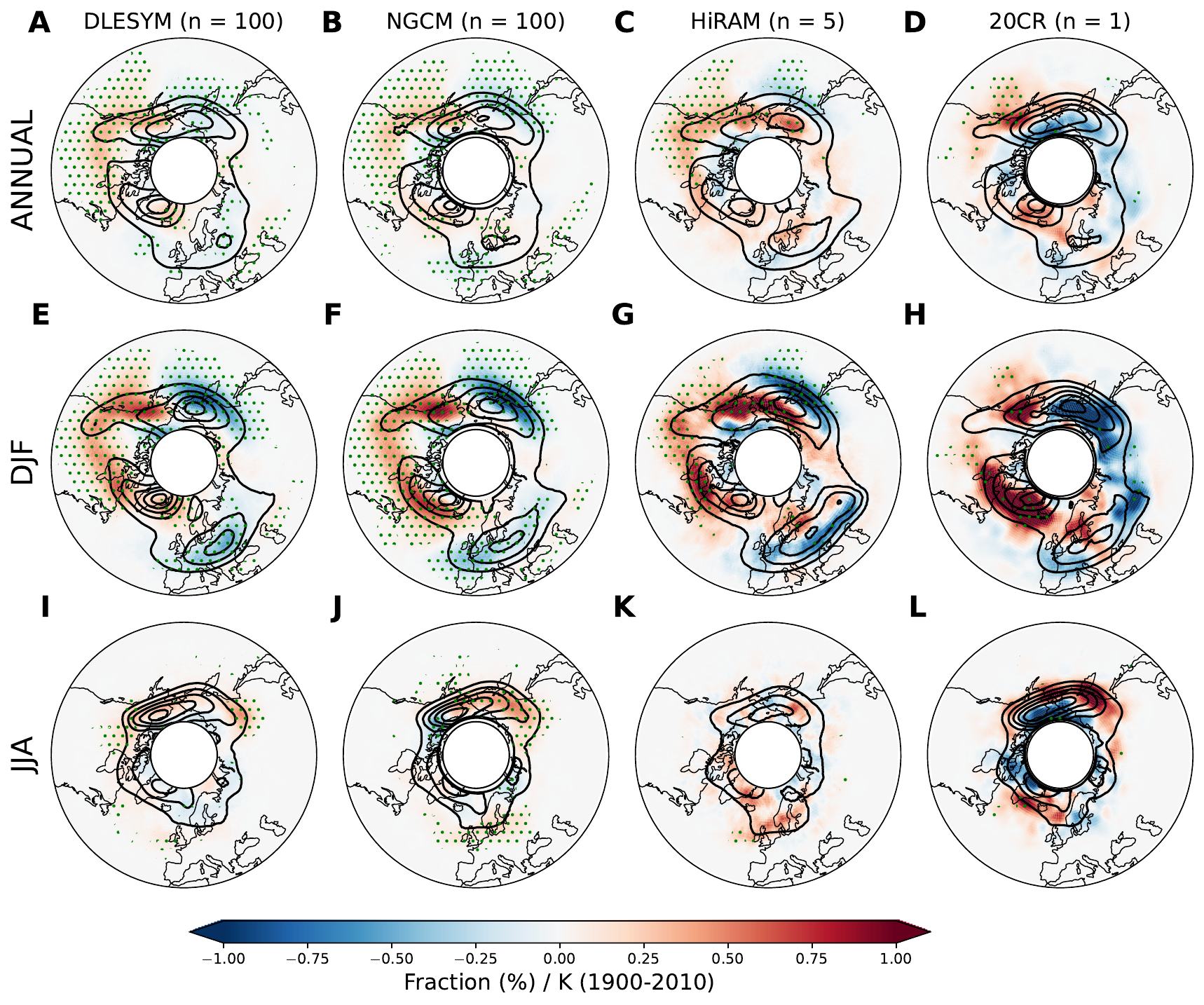}
    \caption{\textbf{Regression of blocking frequency onto Niño3.4 index anomalies}. Spatial patterns of regression slopes between ensemble mean blocking frequency anomalies and Niño3.4 index (1900-2010) for (A-D) annual, (E-H) DJF, and (I-L) JJA seasons. Results are shown for DLESYM (n = 100), NGCM (n = 100), HiRAM (n = 5), and 20CR (n = 1). Shading indicates the percentage change in blocking frequency per Kelvin of Niño3.4 anomaly. Black contours denote climatological blocking frequency patterns. Green stippling marks regions where the regression is statistically significant ($p < 0.01$).}
    \label{sfig:ninoreg}
\end{figure}

\begin{figure}
    \centering
    \includegraphics[width=1\linewidth]{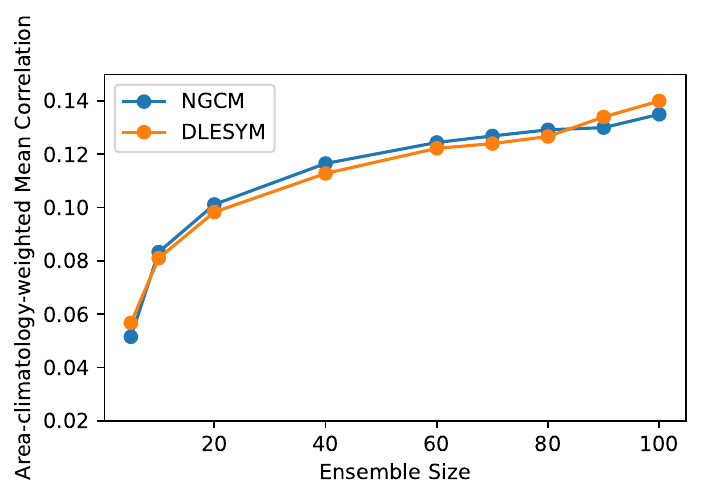}
    \caption{Change in the area-climatology-weighted mean correlation over the Northern Hemisphere as a function of ensemble size, estimated using the bootstrap method.}
    \label{fig:corrwithsize}
\end{figure}

\clearpage
\bibliographystyle{ametsocV6}
\bibliography{ref}  


\end{document}